\documentstyle[twoside,fleqn,espcrc2]{article}

\input epsf
\newcommand{\be}{\begin{equation}} \newcommand{\ee}{\end{equation}}
\newcommand{\bea}{\begin{eqnarray}} \newcommand{\eea}{\end{eqnarray}}
  
 \def\a{\alpha}

\newcommand{\AmS}{{\protect\the\textfont2
  A\kern-.1667em\lower.5ex\hbox{M}\kern-.125emS}}

\title{$\alpha_s(m_Z)$, intermediate scales and 
Fermion masses in supersymmeric theories; two-loop results}

\author{Biswajoy Brahmachari \address{
        International Centre For Theoretical Physics,\\
        34100 Trieste, ITALY.}%
        \thanks{e-mail: biswajoy@ictp.trieste.it}
        }   
       
\begin{document}

\begin{abstract}
One way to generate an intermediate scale being consistent with gauge
coupling unification is to add new Higgs scalars above the intermediate
scale. We classify such scenarios according to their degree of departure
from minimal supersymmetric standard model. Thereafter, we summarize the 
results of a two-loop renormalization group analysis of the gauge and the 
Yukawa sectors of such scenarios, and their sensitivity to the {\it input} 
$\alpha_3(m_Z)$. The presence of an adjoint color octet above the 
intermediate 
scale can raise the gauge coupling unification scale to the $string$ scale, 
acommodate a suitable intermediate scale, provide a left-handed $\tau$-neutrino of 
mass of the order of few electron volts and also have the right prediction of the low 
energy ratio of $m_b /m_\tau$.  
\end{abstract}

\maketitle

There are several physical arguments suggesting that in an unified 
theory like SO(10) there may be an intermediate scale \cite{int1} 
corresponding to a left-right gauge symmetry breaking \cite{intscale} 
somewhere around $10^{11}$ to $10^{12}$ GeV based on
neutrino physics \cite{hdm,numass} as well as strong CP problem
\cite{kim,cpsusy}. In this talk we summarize a two-loop analysis \cite{mar4} of a 
class of intermediate-scale supersymmetric SO(10) models 
\cite{sato,rizzo,leemoh,biswamoha,vissani,bringole} when the strong coupling 
constant $\alpha_3(m_Z)$ in the range 0.110 to 0.130 \cite{pdg} [See Figure (1a)]. 
One-step unification $predicts$ $\a_3(m_Z) \ge 0.126$ including the supersymmetric 
threshold corrections \cite{bagger}.
\begin{figure*}[t]
\begin{tabular}{cc}
\epsfysize=4.5cm \epsfxsize=7cm \hfil \epsfbox{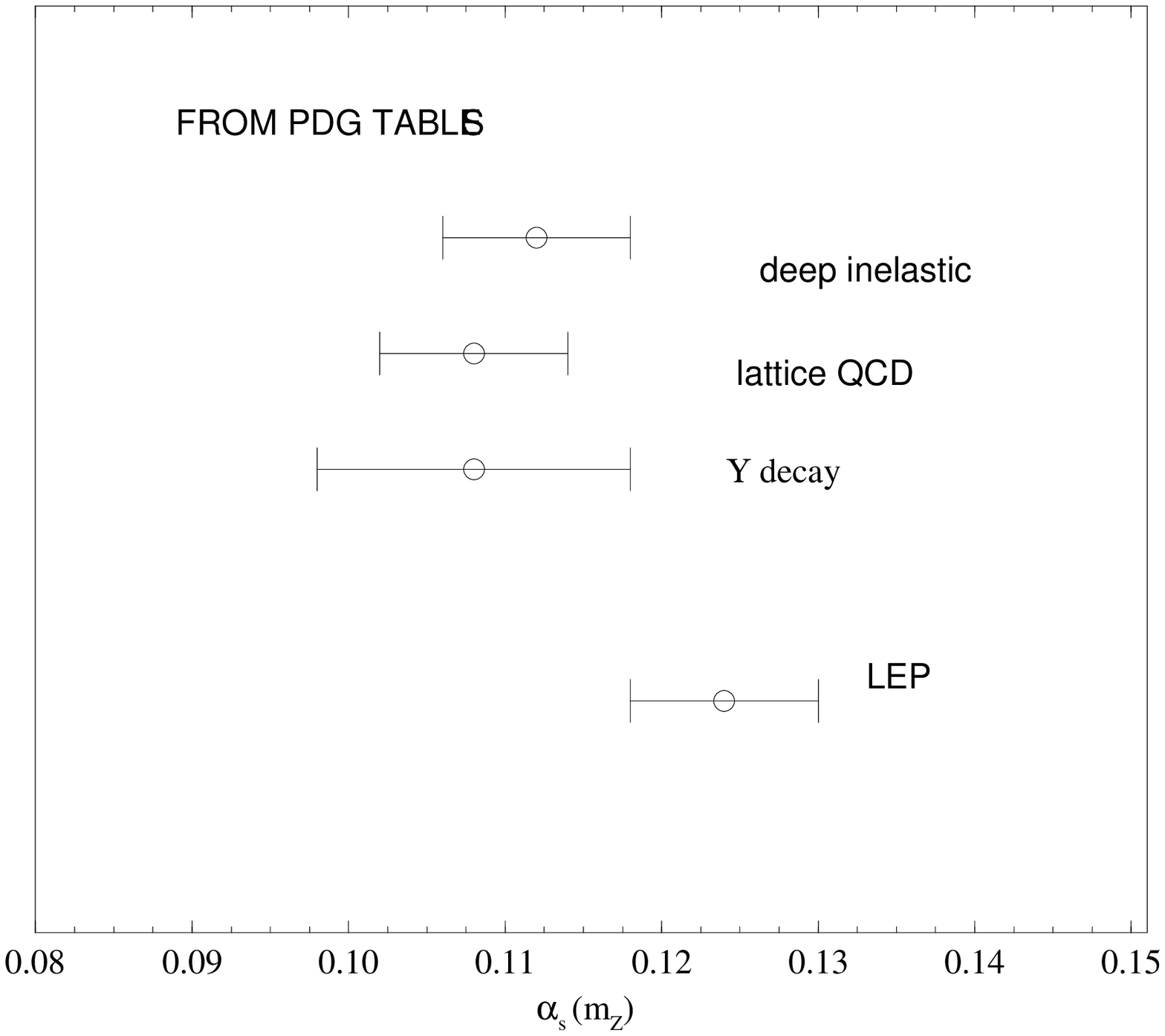} \hfil
&
\epsfysize=4.5cm \epsfxsize=7cm \hfil \epsfbox{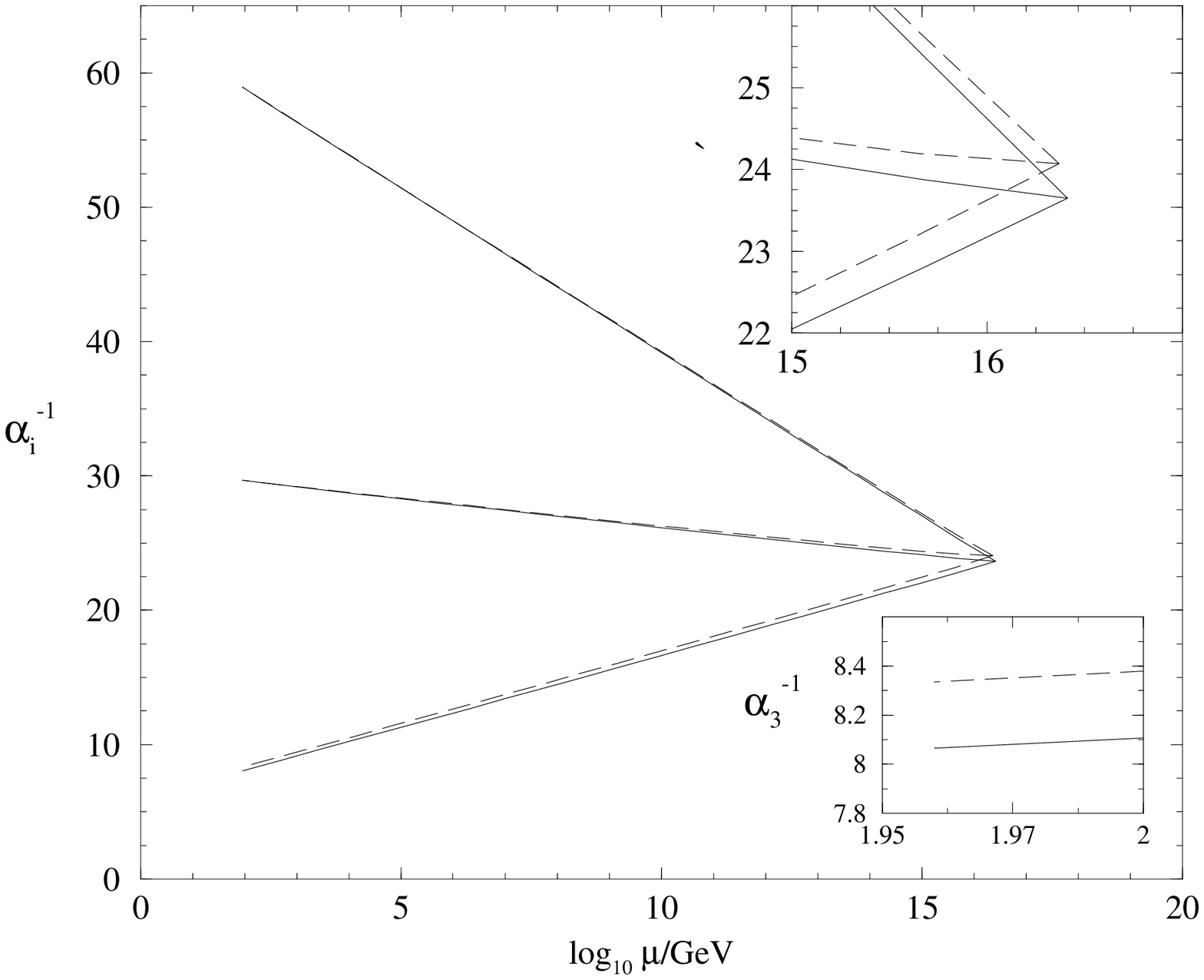} \hfil
\end{tabular}
\caption{ (a) Various measurements of $\alpha_3(m_Z)$ (b) The one-step 
unification 
case. The lower box shows the values of $\alpha_3(m_Z)$ required to achieve 
one-step unification.} 
\end{figure*}

Equivalently, $\a_3(m_Z)$ can also be treated as an $input$. Three inputs in the 
gauge sector namely $\alpha_1(m_Z)$, $\alpha_2(m_Z)$ and $\alpha_3(M_Z)$ can 
determine the three unknown parameters $M_X$, $M_I$ and $\alpha_G(M_X)$. Given
$\alpha_1(m_Z)$ and $\alpha_2(m_Z)$, there exists a value of $\a_3(m_Z)$ for 
which $M_X=M_I$. In one loop approximation this unique value, for which $M_X=M_I$, is 
$\alpha_3(m_Z) \sim 0.1144$. Lee and Mohapatra \cite{leemoh} have 
studied a number of models where it is possible to get an intermediate scale $M_I 
\le M_X $ if $\alpha_3 \ne 0.1144$ and if the scalar structure above $M_I$ is enlarged.
 
The 1-loop renormalization group equation (RGE) of the three couplings introducing 
a general intermediate scale $M_I$ between $M_Z$ and $M_X$ is our starting point. 
We have used $b_i$ to denote the beta function coefficients below the intermediate 
scale and $b_i^\prime$ to denote them above the intermediate scale. 
The relations are,
\begin{eqnarray}
\a_i^{-1}(m_Z)&=& \a_G^{-1} + {b_i \over 2 \pi} \ln{M_I \over M_Z}
+ {b_i ^\prime \over  2 \pi} {M_X \over M_I}. \label{rgint}
\end{eqnarray}
In a combination $\delta$ \cite{bm1} $b_i$ can be eliminated keeping $b_i^\prime$, 
where, 
\be 
\delta =7 \a_3^{-1}(m_Z) - 12 \a_2^{-1}(m_Z) +
5 \a_1^{-1}(m_Z). \label{c2} 
\ee 
Eqn.(\ref{rgint}) and Eqn.(\ref{c2}) together leads to, 
\be 
\delta={ 1 \over 2 \pi} (7 b_3^\prime -12 b_2^\prime +5 b_1^\prime)
\ln {M_X \over M_I} \equiv {\Delta \over 2 \pi} \ln{M_X \over M_I} 
\label{c2b}. 
\ee
We get $\delta=0$ when $\a_3(m_Z)=0.1144$ with $\a_1(m_Z)=0.01696$ and 
$\a_2(m_Z)=0.03371$. For our purposes the intermediate symmetry group is 
$ G_I \equiv SU(3)_c \times SU(2)_L \times SU(2)_R \times U(1)_{(B-L)}$. If we restrict 
ourselves to only those Higgs scalars which can arise from superstring models with 
Kac-Moody levels one or two \cite{lykken}, we can characterize the intermediate 
scale models  by a set of five integers ($n_L,n_R,n_H,n_C,n_d$). Here $n_C$ refers 
to the number of (8,1,1,0), $n_H$ means the number of (1,2,2,0) fields and $n_L$ and
$n_R$ means the number of {(1,2,1,1)+(1,2,1,-1)} and {(1,1,2,1)+(1,1,2,-1)} fields 
under $G_I$ and $n_d$ refers to the number of pairs of Higgs doublets below 
the scale $M_I$ where the gauge symmetries are that of MSSM. Inserting the relevant 
beta function coefficients \cite{mar4} and using $0.110 \le \alpha_3(m_Z) 
\le 0.130$ we get,
\be
-10 \le (-9 + 21 n_c - 9 n_H + 6 n_R - 9 n_L) \le 6.
\label{ineq} 
\ee
Table \ref{table2} catalogs these models in the decreasing 
order of minimality. In scenario IX the theory below the scale $M_I$ has four Higgs 
doublets ($n_d=2$). 

Now we are in a position to proceed to a full two-loop analysis. 
\begin{table*}[hbt]
\setlength{\tabcolsep}{1.5pc}
\newlength{\digitwidth} \settowidth{\digitwidth}{\rm 0}
\catcode`?=\active \def?{\kern\digitwidth}
\caption{The minimal models which satisfy the condition in Eqn.(4). 
When the quantity $\Delta$ is positive (negative) the model gives rise 
to an intermediate scale for the lower (higher) values of 
$\a_3(m_Z)$ than the one step unification case for which 
$\a_3(m_Z)=0.1144$ at the one-loop level. In the case $\Delta=0$ the 
intermediate scale is unconstrained at the one-loop level.} 
\label{table2}
\begin{tabular*}{\textwidth}{@{}l@{\extracolsep{\fill}}||c| c c c c c||c}
\hline
Model &$n_L$& $n_R$& $n_H$& $n_c$ & $n_d$ & $\Delta$ \\ 
\hline 
I & 0 & 2 & 1 & 0 & 1 & -6\\ 
II & 0 & 3 & 1 & 0& 1 &0\\
III & 0 & 4 & 1 & 0& 1 &6\\  
IV & 0 & 3 & 2 & 0 & 1 &-9\\ 
V & 0 & 4 & 2 & 0 & 1 &-3\\
VI & 0 & 5 & 2 & 0 & 1 & 3\\ 
VII & 1 & 5 & 2 & 0 & 1 &-6\\
\hline 
VIII& 1 & 1 & 1 & 1 &1 &0\\ 
\hline
IX&0&3&2&1&2&--\\
\hline
\end{tabular*}
\end{table*}
The Yukawa couplings can be defined by the superpotential invariant
under the intermediate symmetry. The variation of the unification scales with that 
of $\a_3(M_Z)$ have been plotted \cite{mar4} in Figure (2a). The predictions of the 
intermediate scale have been plotted in Figure (2b) and that of the unification 
coupling $\alpha_G(M_X)$ have been plotted in Figure (2c) for various
models. The grand desert case can be recovered from the meeting point
of all the curves in Figure (2b) that is when the intermediate scale is
equal to the GUT scale. In model VI the unification scale becomes low in the low 
$\a_3$ region; for model V the same thing happens for high $\a_3(M_Z)$ region.  As 
the dimension five proton decay can be suppressed in the SO(10) models by some 
additional mechanism \cite{babr}, we plot the dominant dimension six decay mode in 
Figure (2d). 
\begin{figure*}[t]
\begin{tabular}{cc}
\epsfysize=4.5cm \epsfxsize=7cm \hfil \epsfbox{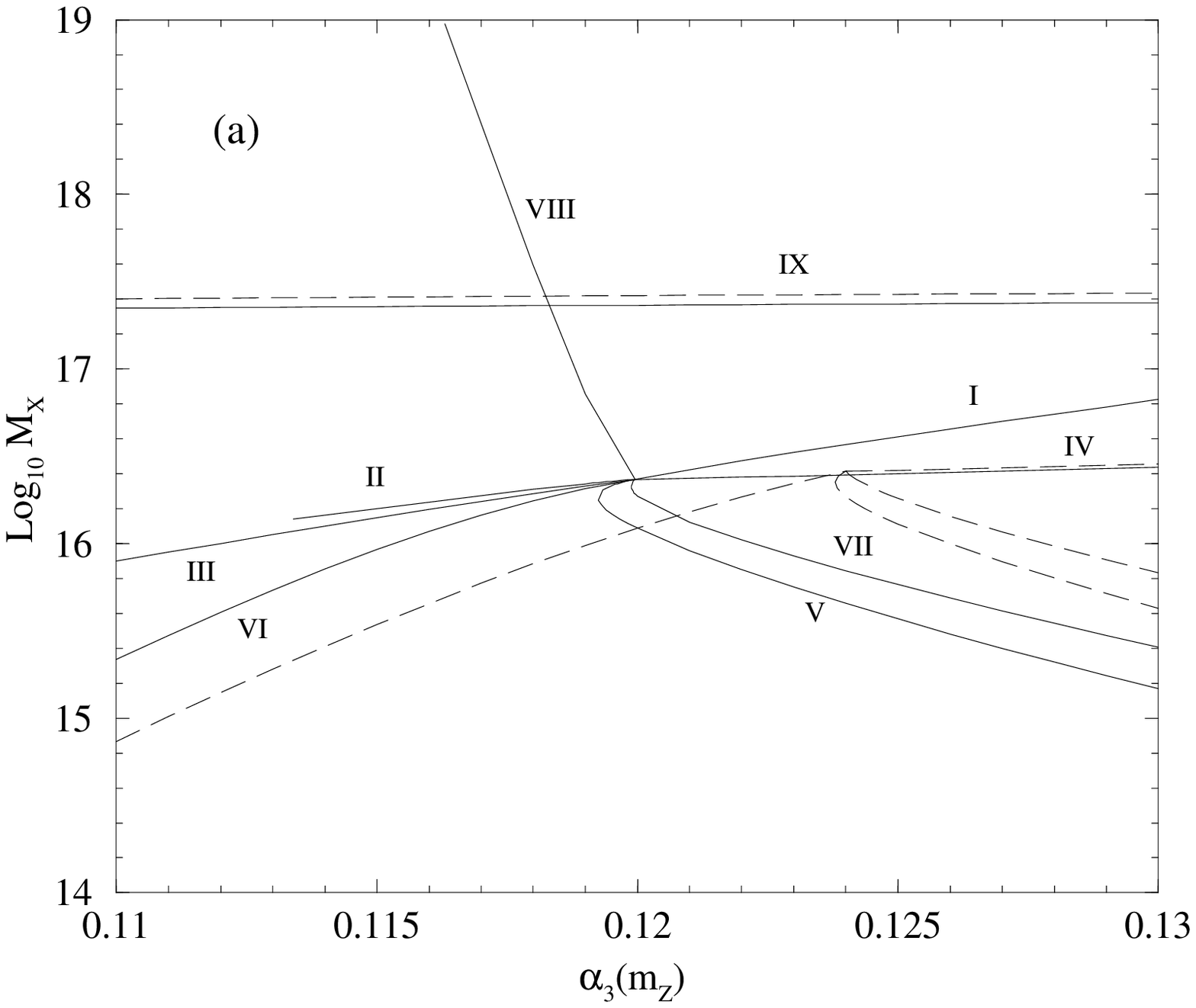} \hfil
& \epsfysize=4.5cm \epsfxsize=7cm \hfil \epsfbox{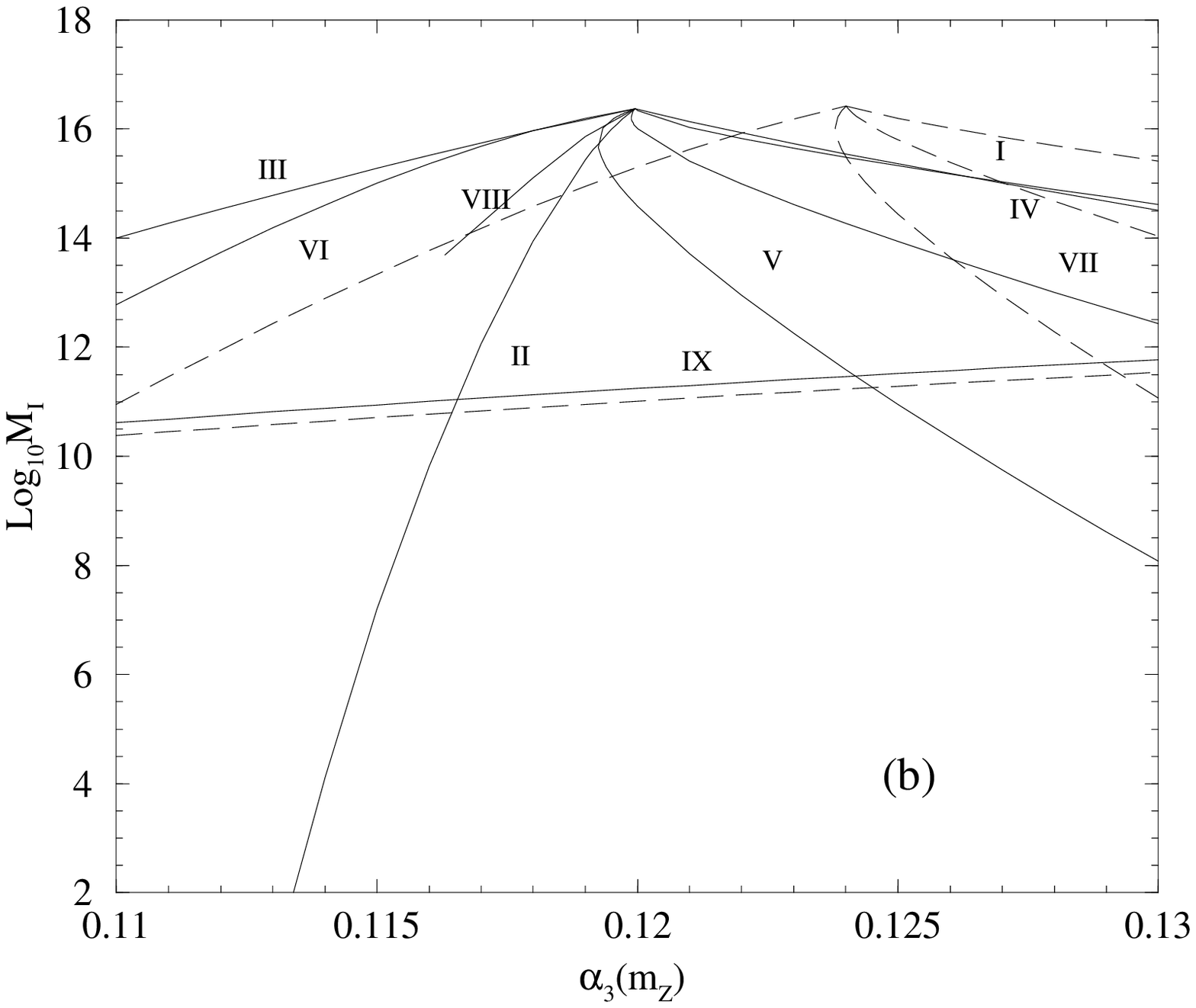} \hfil
\\ \epsfysize=4.5cm \epsfxsize=7cm \hfil \epsfbox{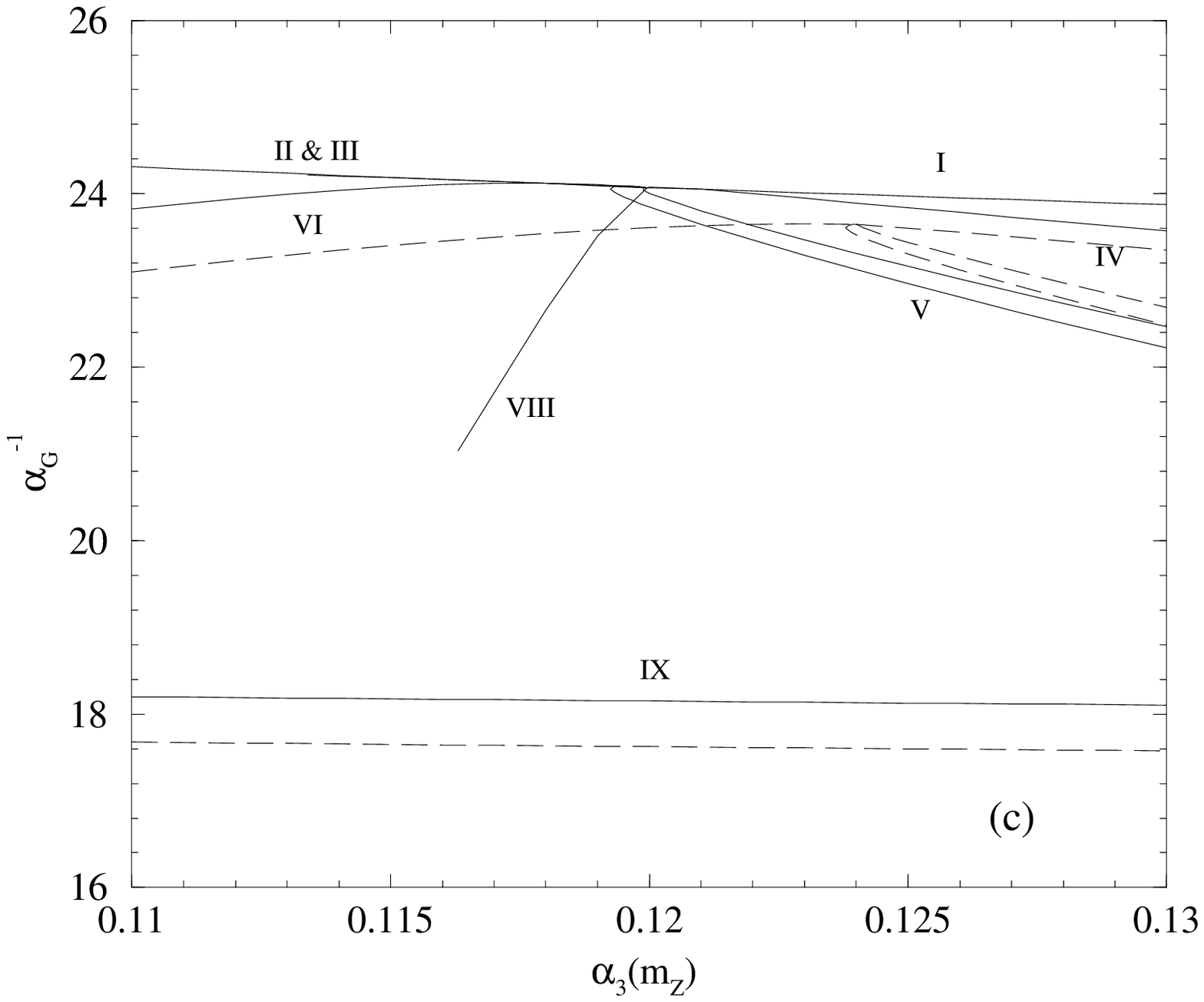} \hfil
& \epsfysize=4.5cm \epsfxsize=7cm \hfil \epsfbox{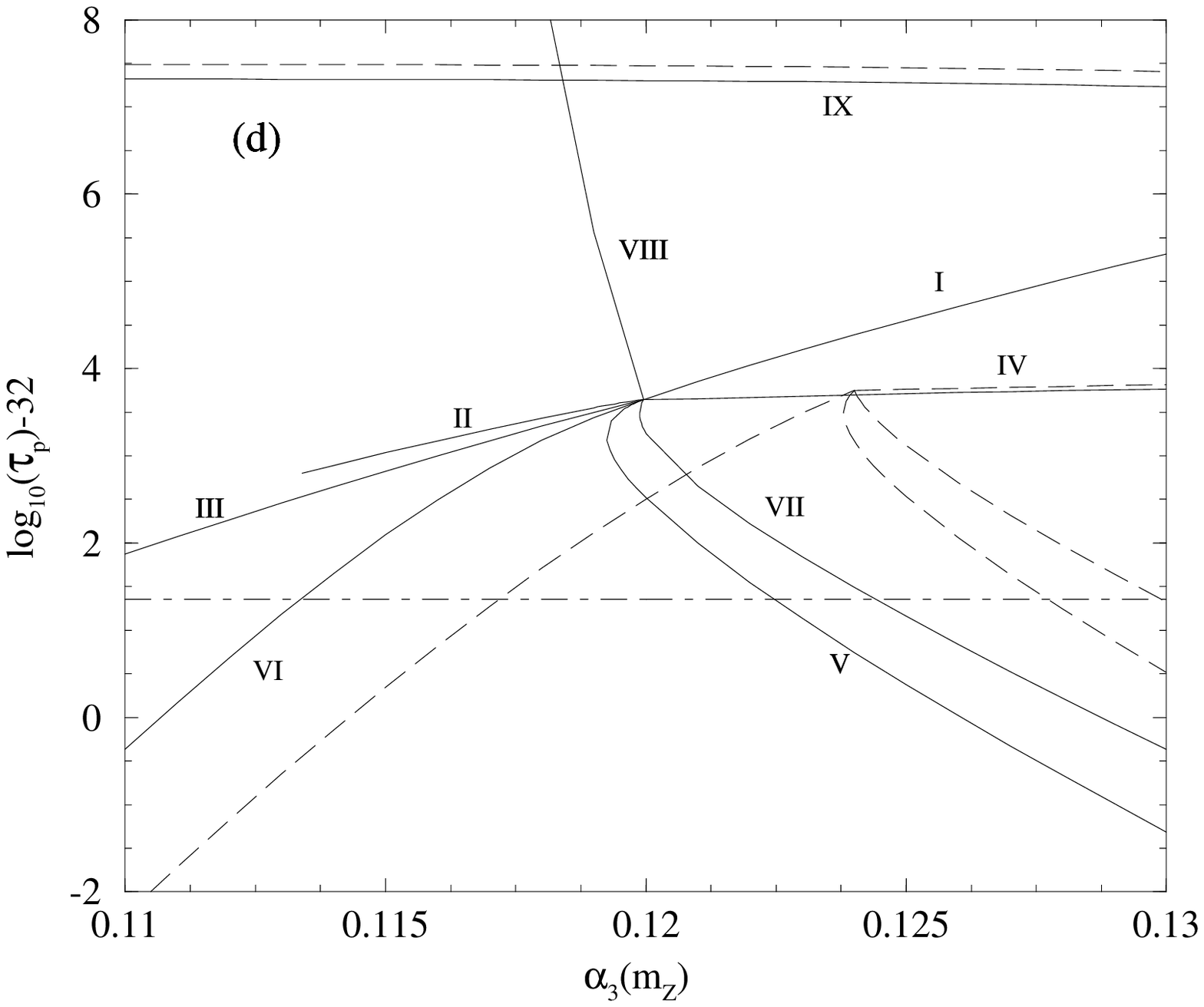} \hfil
\end{tabular}
\caption{ Predictions for (a) Unification scale, (b) Intermediate
scale, (c) Unification gauge coupling, and (d) Proton life--time, for
the models listed in Table I. Solid lines denote high $\tan \beta$
($Y_1(M_X)=Y_2(M_X)=1$; dashed lines denote the low $\tan \beta$
regime ($Y_1(M_X)=1$, $Y_2(M_X)=10^{-4}$). In Figure (d) the dotted
line is the experimental limit $\tau_p=5.5 \times 10^{32}\,yr$.}
\end{figure*} 
Near the GUT scale the Yukawa 
couplings are large and they fall quickly below the GUT scale. The Yukawa 
couplings tend to pull the individual lines towards the low $\a_3$ region 
wheres in the high $\a_3$ models (like V and VII) the gauge interactions have 
exactly the reverse effect. This causes the curvature in the graphs near 
the GUT scale which is purely a two-loop effect. 
\begin{figure*}[t]                                                        
\begin{tabular}{cc} 
\epsfysize=4.5cm \epsfxsize=7cm \hfil \epsfbox{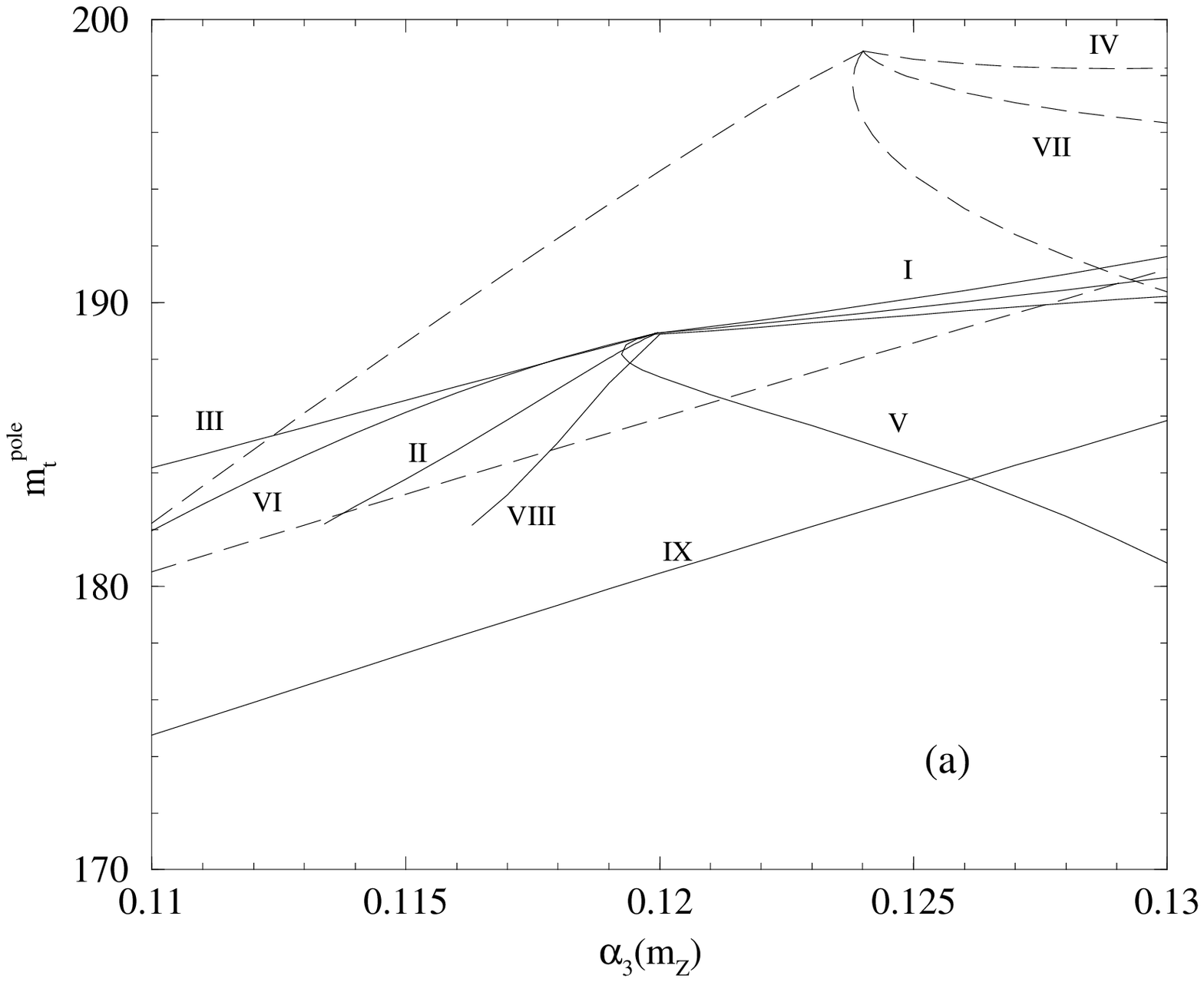} \hfil 
&
\epsfysize=4.5cm \epsfxsize=7cm \hfil \epsfbox{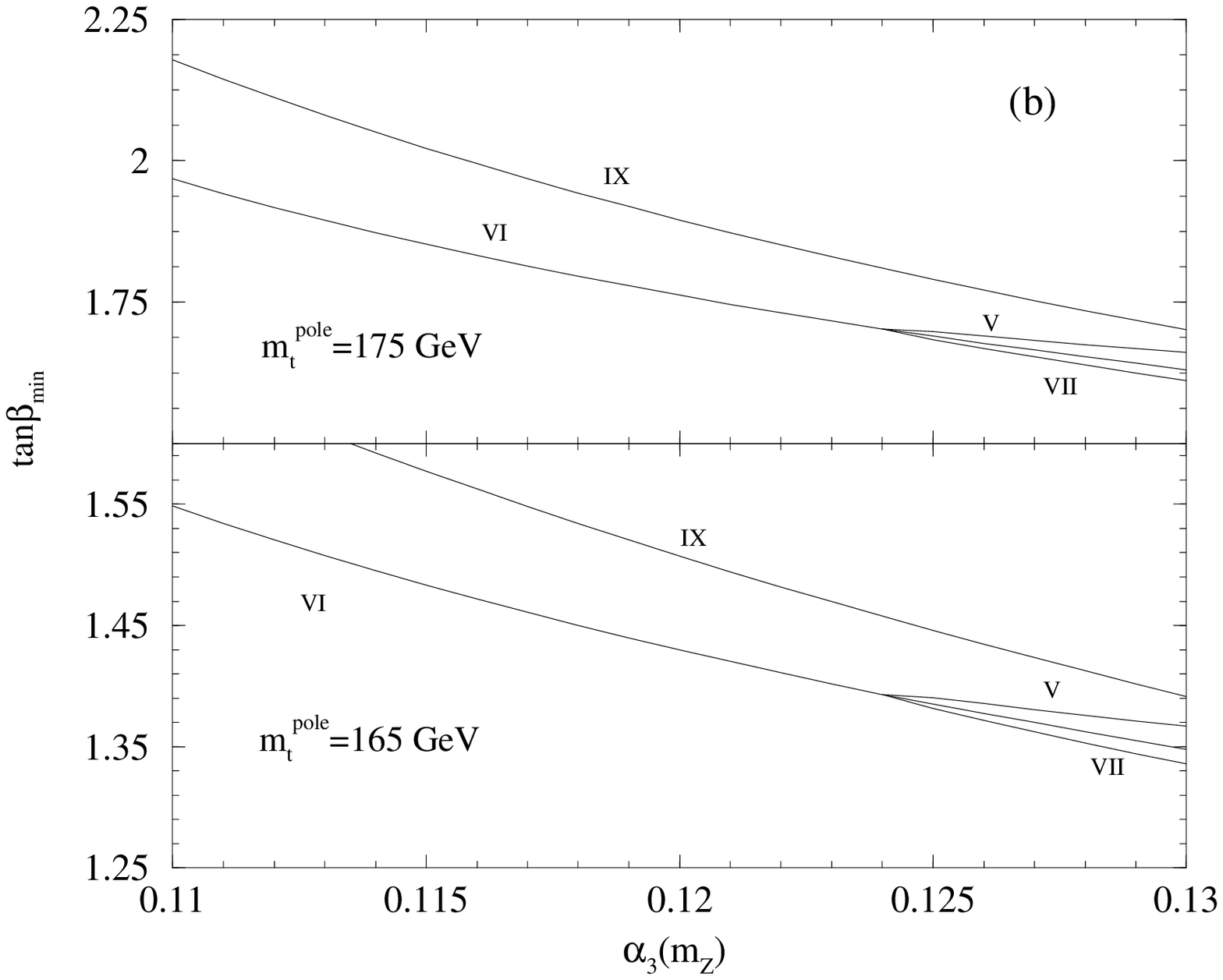} \hfil
\\
\epsfysize=4.5cm \epsfxsize=7cm \hfil \epsfbox{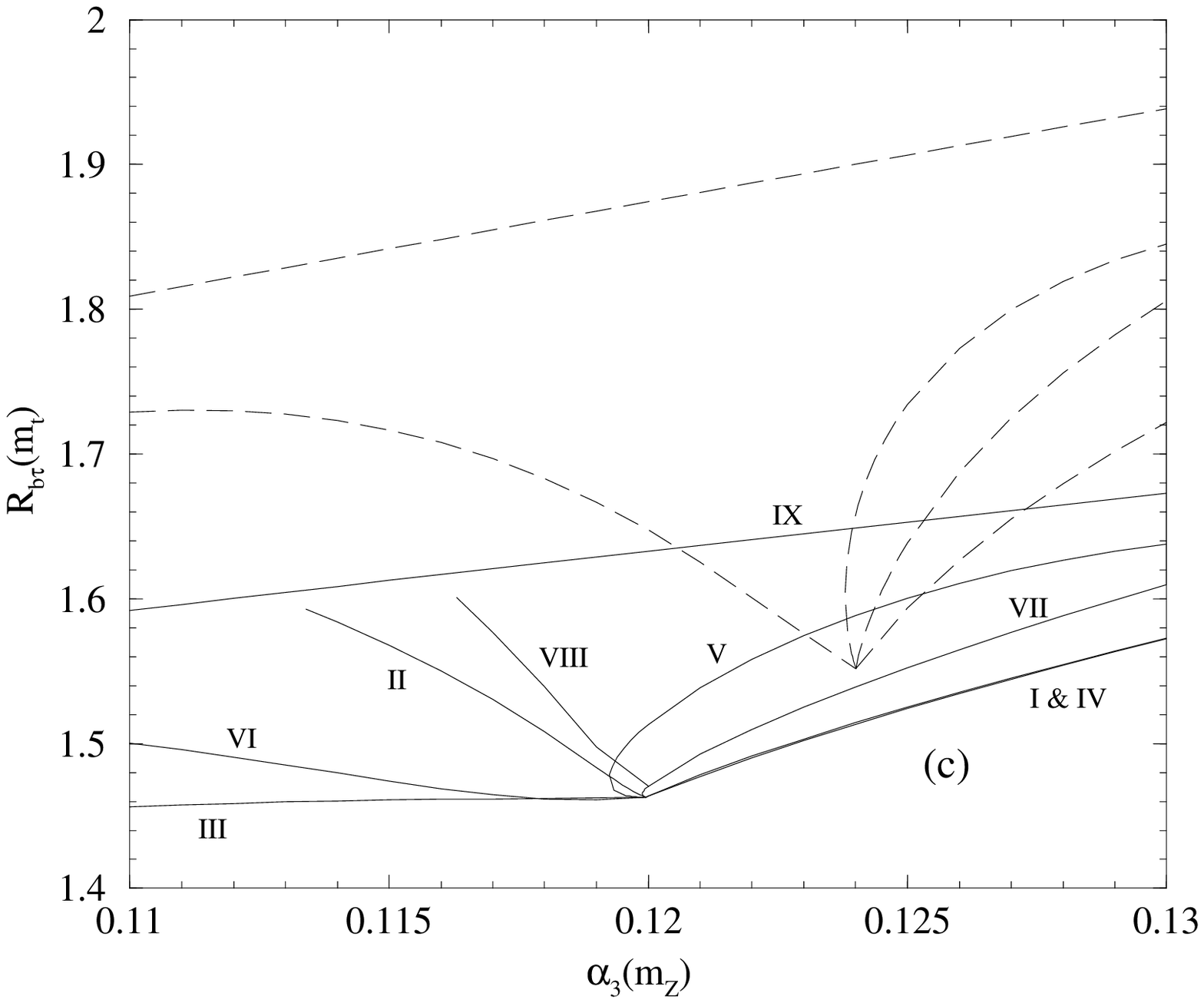} \hfil
&
\epsfysize=4.5cm \epsfxsize=7cm \hfil \epsfbox{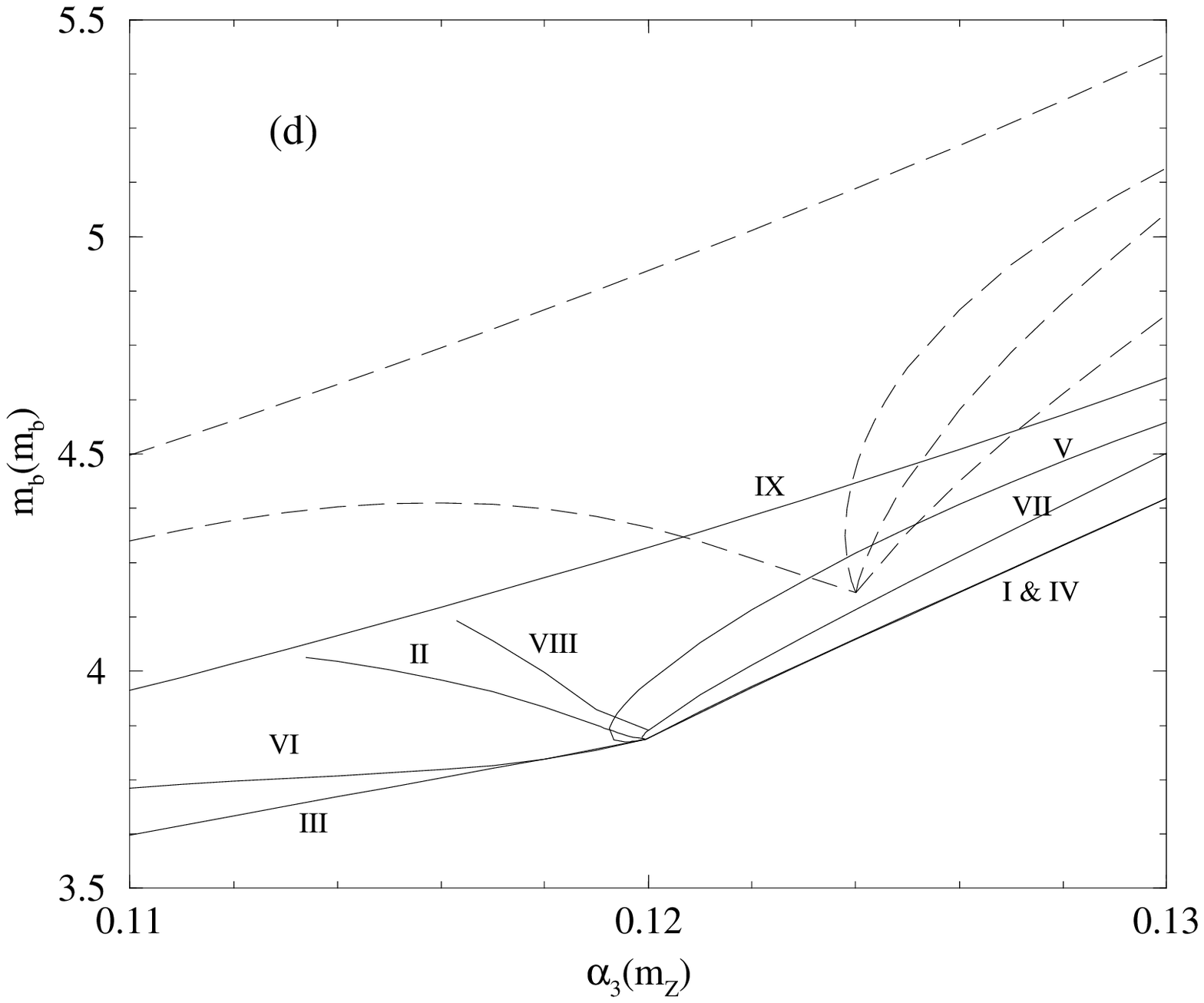} \hfil
\end{tabular}                                                                 
\caption{ Predictions of (a) Pole mass of the top quark, (b) Lower
bound on $\tan \beta$, (c) $R_{b\tau}$ at $m_t$, (d) Running mass of
the b quark. Solid lines and dashed lines are as in Fig. (1).}  
\end{figure*} 
Using $m_\tau(m_\tau)$ GeV, we calculate the value of $\tan \beta$ at low energy. 
Once the value of $\tan \beta$ is known unique predictions for $m_t(m_t)$ and 
$m_b(m_b)$ follows. The pole mass \cite{mtpole} has been calculated from the 
running mass for each value of $\alpha_3(m_Z)$. The predictions of 
$m^{pole}_t$ is plotted in Figure (3a). The value of $R_b(m_t) \equiv {m_b(m_t) \over
m_\tau(m_t)}$ are plotted in Figure (3c), and when the prediction of
$m_b(m_t)$ is extrapolated to the mass scale of the bottom quark we
get Figure (3d). The lower bounds on $\tan \beta (\sin \beta)$
follows directly from an upper bound on the top quark Yukawa coupling. These
bounds have been plotted in Figure (3b). In our models the intermediate $B-L$ 
symmetry is broken by the Higgs scalars $16+\overline{16}$ fields of SO(10).  We 
will consider two different scenarios by which Majorana mass of the right handed
neutrino can be generated. (a) Using a higher dimensional operator of the 
form ${h \over M_X} 16_F 16_F 16_H 16_H $ written in terms of SO(10)
representations. The subscripts $F$ and $H$ mean fermions and scalars
respectively. When $16_H$ gets a VEV a large Majorana mass of the
order $h~v^2_R/M_X$ is generated. (b) Introduction of additional singlets 
to have a generalized see-saw mechanism \cite{leemoh,moha86}. 
\begin{figure*}[t]                                                        
\begin{tabular}{cc} 
\epsfysize=4.5cm \epsfxsize=7cm \hfil \epsfbox{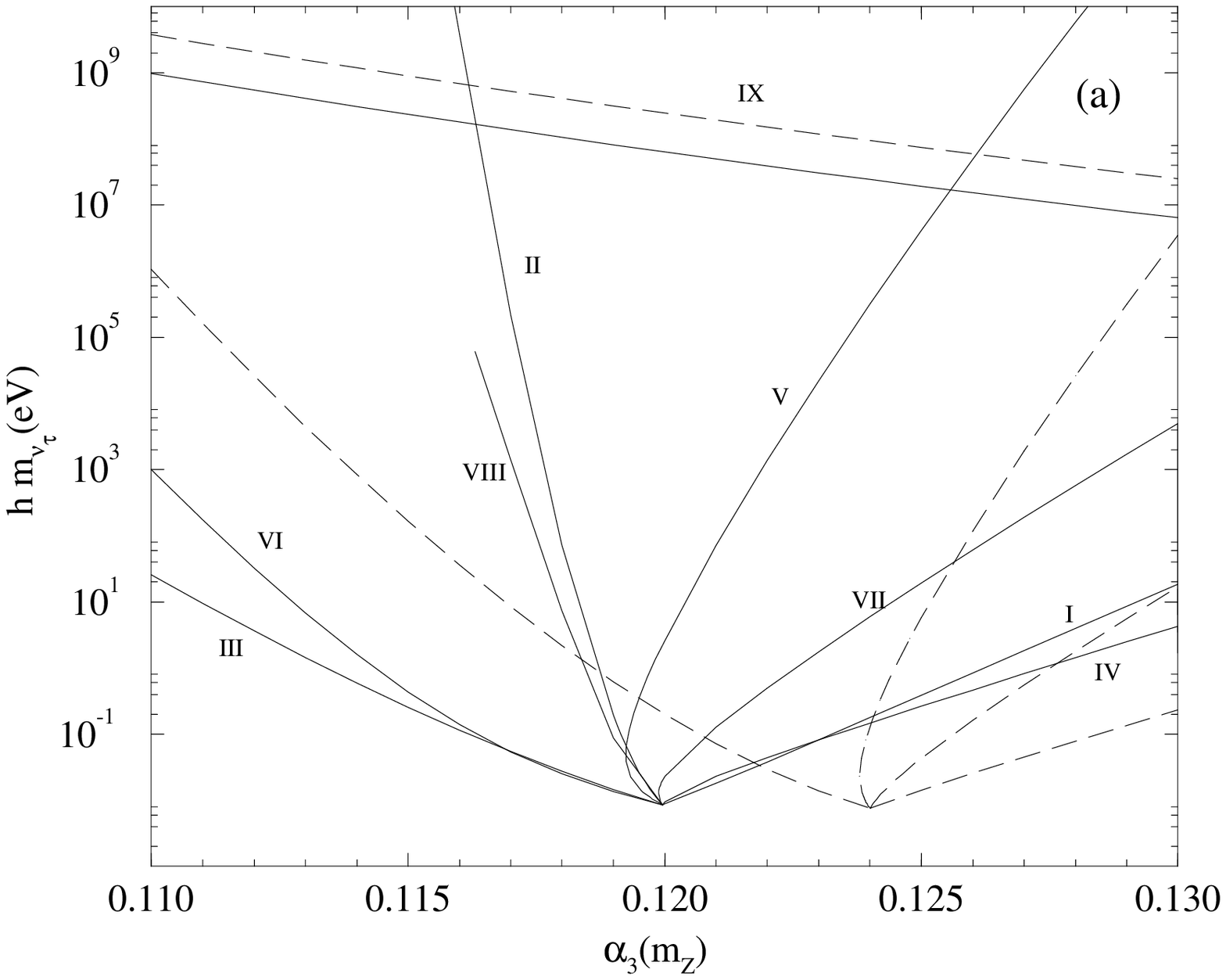} \hfil
&
\epsfysize=4.5cm \epsfxsize=7cm \hfil \epsfbox{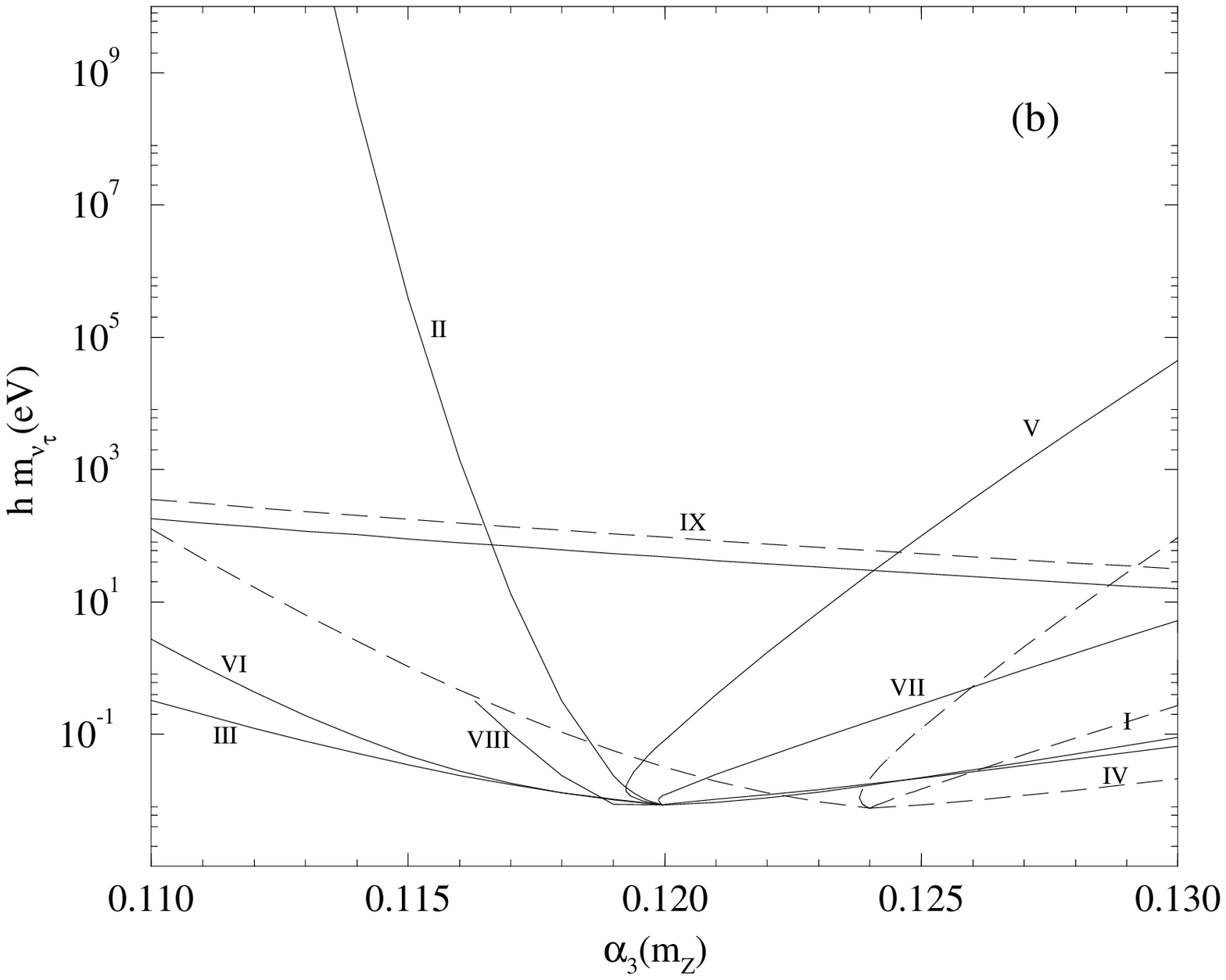} \hfil
\end{tabular} 
\caption{ Predictions of the left--handed neutrino mass by see--saw
mechanism by the two scenarios (a) and (b). Solid and dashed lines are
as before.}
\end{figure*}
In Figure (4a) and Figure (4b) we have plotted the left handed neutrino
masses in scenarios (a) and (b) modulo the unknown Yukawa couplings $h$ and 
$h^\prime$ in various models [see Ref\cite{mar4}]as a function of 
$\a_3(m_Z)$. 

We know that a tau neutrino mass of the order of a few electron volts
is preferable if neutrino is to be a candidate for the Hot Dark Matter
(HDM). We see that in scenario (a) a tau neutrino in the range of
1-10 eV can be achieved in all the models depending on the value of
$\a_3(m_Z)$. On the other hand, scenario (b) can predict a tau
neutrino mass in the 1-10 eV range for models V, II and VI. 

To conclude, we have summarized a two-loop RGE analysis of the
gauge and Yukawa couplings in a class os SUSY unified theories
with intermediate scales. The presence of a color octet (model VIII) above the
intermediate scale can push the unification scale to the $string$ scale for
$\alpha_3$ in the range $0.118-0.119$. In this scenario the bottom quark 
mass 
prediction is attractive and the $\tau$ neutrino has a suitable mass to 
become a 
candidate of hot dark matter. The equality of the left and right handed gauge 
couplings also remain preserved above the intermediate scale in this 
scenario as $n_L=n_R$.

I thank my collaborators M. Bastero-Gil and R. N. Mohapatra.

\end{document}